\documentclass[aps,prl,showpacs]{revtex4}
\usepackage{graphicx}
\usepackage{amsfonts,amssymb,amsmath}
\usepackage{bm}

\begin{document}

\title{Solution of two mode bosonic Hamiltonians and related physical systems}
\date{\today}
 
\author{Ramazan Ko\c{c}}
\email{koc@gantep.edu.tr}
%\homepage{www.mrl.ucsb.edu/~ederer} 
\affiliation{Department of Physics, Faculty of Engineering 
University of Gaziantep,  27310 Gaziantep, Turkey}
\author{Hayriye T\"{u}t\"{u}nc\"{u}ler}
\email{tutunculer@gantep.edu.tr}
\affiliation{Department of Physics, Faculty of Engineering 
University of Gaziantep,  27310 Gaziantep, Turkey}
\author{Mehmet Koca}
\email{kocam@squ.edu.om}
\affiliation{Department of Physics, College of Science,
Sultan Qaboos University, PO Box 36  \\
Al-Khod 123, Sultanete of Oman}

\begin{abstract}
 We have constructed the quasi-exactly-solvable two-mode bosonic
 realizations of $su(2)$ and $su(1,1)$ algebra. We derive the relations
 leading to the conditions for quasi-exact-solvabilty of two-boson
 Hamiltonians by determining a general procedure which maps the Schwinger
 representations of the $su(2)$ and $su(1,1)$ algebras to the Gelfand-Dyson
 representations respectively. This mapping allows us to study nonlinear
 quantum optical systems in the framework of quasi-exact-solvabilty. Our
 approach also leads to a simple construction of special functions of two
 variables which are the most appropriate functions to study
quasi-probabilities in quantum optics.
\end{abstract}
\maketitle

\section{Introduction}

The exact solutions of the multi-bosons systems play an important role in
quantum mechanics. The Hamiltonians are employed in quantum optics most
often involving multi-bosons. During last decade a great attention has been
paid to examine different quantum optical models with Hamiltonians given by
multi-bosons\cite{qu,kara1,kara2,alv,klim}. The quantum optical systems
whose Hamiltonians expressed by Casimir operator of the symmetry group can
exactly be solved\cite{gursey}. The quantum optical systems with
Hamiltonians including nonlinear functions in Lie algebra generators can be
analyzed by using mainly numerical methods, because Lie algebraic techniques
are non-efficient for such systems and most of the other analytical
techniques require in general tedious calculations.

In quantum mechanics there exist physical systems whose finite part of
spectrum can be exactly obtained in closed forms. They are known as
quasi-exactly-solvable(QES) systems\cite{turb1,gonz,ushw1,bender,shifman,roy}%
. Dating back over fifteen years there has been a great deal of interest in
QES systems. The one dimensional QES systems based on $sl_{2}$-algebra have
been classified by Turbiner\cite{turb1}. The usual approach to the analysis
of the one dimensional QES systems is to express the Hamiltonian of the
corresponding physical system as linear and bilinear combination of the
generators of the $sl_{2}$- algebra. The necessary conditions for the
normalizabilty of the wavefunction of the QES systems based on $sl_{2}$-
algebra were completely determined in\cite{gonz}. The QES models, either in
the forms of differential equations or of the single boson Hamiltonian, have
been extended by Dolya and Zaslavskii\cite{dolya1,dolya2}.

The aim of this paper is to extend the QES systems to the multi-boson
systems and to obtain the solution of the multi-boson Hamiltonians, in
particular, Hamiltonians of the nonlinear quantum optical systems. The
formalism presented here also leads to the construction of special functions
with two variables which are appropriate to study many quantum optical
problems and quasi-probabilty\cite{alfred1,alfred2}. We think that the QES
bosonic systems deserve special treatments for they have many applications
in physics. The suggested approach can be extended to the matrix
multi-bosonic systems.

In the Bosonic Hamiltonian, it is more convenient to use the bosonic
representations of the $su(2)$ and $su(1,1)$ generators and determine the
conditions of quasi-exact solvability. The single boson realization of the $%
su(1,1)$ algebra has been studied in\cite{dolya1,dolya2}. In this work we
follow a different strategy to obtain the conditions for QES of the
two-bosonic systems.

The algebras of the groups $SU(2)$ and $SU(1,1)$ have useful applications in
quantum physics\cite{wyb,alhassid1,alhassid2}. They are used to generate the
energy spectra while the representation matrices of the group could be used
to calculate time dependent excitations of the bound states and the
scattering states respectively\cite{gursey}. As usual the group elements and
their associated algebras can be expressed in terms of bosonic operators.

This paper is organized as follows. The constructions of the two-mode
bosonic realization of the $su(2)$ and $su(1,1)$ algebra are briefly
reviewed in section 2. In section 3, we propose the transformation of the
boson creation and annihilation operators leads to the two class $su(2)$ and
two class $su(1,1)$ one dimensional realizations that is useful to study QES
systems. The one dimensional realizations of the $su(2)$ and $su(1,1)$
algebras are presented by similarity transformation section 4. We also
discuss the some applications of our approach. We demonstrate the procedure
presented here can be used to solve a rich family of physical systems. In
particular, the Karassiov-Klimov Hamiltonian, Hamiltonians of the second and
third generation harmonic oscillator systems, Hamiltonian of the quantal
systems under thermal effect have been solved in section 5. We also
demonstrate the relations between QES one-dimensional Schr\"{o}dinger
equation and two-mode bosonic Hamiltonians leads to the construction of the
two dimensional differential equations of the special functions, in this
section. The paper ends with a brief discussion and conclusion.

\section{Construction of two-boson Hamiltonians: $su(2)$ and $su(1,1)$
algebras}

A convenient way\cite{gursey} to construct a spectrum generating algebra for
systems with a finite number of bound states is by introducing a set of
boson creation and annihilation operators. We introduce two boson operator, $%
a_{1}$ and $a_{2}$, obey the usual commutation relations

\begin{equation}
\left[ a_{1},a_{1}\right] =\left[ a_{1},a_{2}^{+}\right] =\left[
a_{2},a_{2}^{+}\right] =0,\quad \left[ a_{1},a_{1}^{+}\right] =\left[
a_{2},a_{2}^{+}\right] =1.  \label{eq:1}
\end{equation}

The bilinear combinations $a_{1}^{+}a_{1},a_{1}^{+}a_{2},a_{2}^{+}a_{1}$ and 
$a_{2}^{+}a_{2}$ generates the group $U(2)$ and $a_{1}^{+}a_{1}$, $%
a_{1}a_{2} $, $a_{1}^{+}a_{2}^{+}$ and $a_{2}^{+}a_{2}$ generates the group $%
SU(1,1)$. Let us start by introducing three generators of $SU(2)$,

\begin{equation}
J_{+}=a_{1}^{+}a_{2},\quad J_{-}=a_{2}^{+}a_{1},\quad J_{0}=\frac{1}{2}%
\left( a_{1}^{+}a_{1}-a_{2}^{+}a_{2}\right) .  \label{eq:2}
\end{equation}%
These are the Schwinger representation of $SU(2)$ algebra and they satisfy
the commutation relations

\begin{equation}
\left[ J_{+},J_{-}\right] =2J_{0}\quad \left[ J_{0},J_{\pm }\right] =\pm
J_{\pm }  \label{eq:3}
\end{equation}%
The fourth generator is the total boson number operator

\begin{equation}
N=a_{1}^{+}a_{1}+a_{2}^{+}a_{2}  \label{eq:4}
\end{equation}%
which commutes with the $SU(2)$ generators. The Casimir operator of this
structure is given by

\begin{equation}
J=J_{-}J_{+}+J_{0}(J_{0}+1)=\frac{1}{4}N(N+2).  \label{eq:5}
\end{equation}%
If we denote the eigenvalues of the operator $J$ by

\begin{equation}
J=j(j+1)  \label{eq:6}
\end{equation}%
the irreducible representations of $SU(2)$ are characterized by the total
boson number

\begin{equation}
j=N/2  \label{eq:7}
\end{equation}%
where $N=0,1,2,...$ .

The Schwinger representation of $su(1,1)$ algebra can be constructed by
considering bosonic realizations of the generators:

\begin{equation}
K_{+}=a_{1}^{+}a_{2}^{+},\quad K_{-}=a_{2}a_{1},\quad K_{0}=\frac{1}{2}%
\left( a_{1}^{+}a_{1}+a_{2}^{+}a_{2}+1\right) .  \label{eq:8}
\end{equation}
They satisfy the commutation relations

\begin{equation}
\left[ K_{+},K_{-}\right] =-2K_{0}\quad \left[ K_{0},K_{\pm }\right] =\pm
K_{\pm }.  \label{eq:9}
\end{equation}%
The total boson number operator of this algebra is given by

\begin{equation}
L=a_{1}^{+}a_{1}-a_{2}^{+}a_{2},  \label{eq:10}
\end{equation}%
and it commutes with the generators of the $su(1,1)$ algebra. The Casimir
operator of this structure can be expressed in terms of number operator,
such that

\begin{equation}
C=-J_{-}J_{+}+J_{0}(J_{0}+1)=\frac{1}{4}(L+1)(L-1).  \label{eq:11}
\end{equation}%
It is obvious that number operator and Casimir operator commute. If the
eigenvalues of the operator $C$ is denoted by

\begin{equation}
C=\ell (\ell +1)  \label{eq:12}
\end{equation}%
the irreducible representations of $su(1,1)$ algebra are characterized by a
total boson number

\begin{equation}
L=-(2\ell +1).  \label{eq:13}
\end{equation}

For the exactly solvable case the Hamiltonian have been written in terms of
Casimir invariants of the algebra. In this case the eigenvalues and
eigenfunction of the Hamiltonian can be obtained in the closed form. Our
task is now to obtain the conditions for QES of the $su(2)$ and $su(1,1)$
algebra given in (\ref{eq:2}) and (\ref{eq:8}). The conditions can be
obtained by connecting two-mode bosonic realizations of the $su(2)$ and $%
su(1,1)$ algebra to the $sl(2,R)$ and $sl_{2}(R)$ algebra, respectively.
Since Casimir operator $J$\ commutes with $N$, and $C$ commute with $L$ the
linear combinations of the generators of each algebra can be diagonalized
within the representation $[N]$ and $[L]$, respectively. The abstract boson
algebra can be associated with the exactly soluble Schr\"{o}dinger equations
by using the differential operator realizations of boson operators. This
connection opens the way to an algebraic treatment of a large class of
potentials of practical interest.

\section{Transformation of the bosonic operators}

In the previous section we have summarized the construction of the two-mode
bosonic realizations of the $su(2)$ and $su(1,1)$ algebras. In this section
we develop a procedure to transform the generators in the Schwinger
representation to a more suitable representation to determine its relation
with the QES systems. This can be done by various methods. Here we follow a
different strategy to obtain the connection between the two-mode bosonic
systems and QES systems. Let us introduce the following similarity
transformation induced by the operator

\begin{equation}
S=(a_{2}^{+})^{\alpha a_{1}^{+}a_{1}}  \label{eq:14}
\end{equation}%
where $\alpha $ is a constant and in order to obtain Gelfand-Dyson
representations of the $su(2)$ and $su(1,1)$ algebra it will be constrained
to $\pm 1$. In general, it will be shown later, it is not necessary to set $%
\alpha =\pm 1,$ to construct QES systems. This property allow us to study a
wide range of physical systems. The operator $S$ acts on the state $\left|
n_{1},n_{2}\right\rangle $ as follows,

\begin{equation}
S\left| n_{1},n_{2}\right\rangle =(a_{2}^{+})^{\alpha n_{1}}\left|
n_{1},n_{2}\right\rangle =\sqrt{\frac{n_{2}!}{(n_{2}+\alpha n_{1})!}}\left|
n_{1},n_{2}+\alpha n_{1}\right\rangle .  \label{eq:15}
\end{equation}%
Since $a_{1}$ and $a_{2}$ commute, the transformation of $a_{1}$ and $%
a_{1}^{+}$ under $S$ can be obtained by writing $a_{2}^{+}=e^{b}$, with $%
[a_{1},b]=[a_{1}^{+},b]=0$,

\begin{eqnarray}
Sa_{1}^{+}S^{-1} &=&e^{\alpha ba_{1}^{+}a_{1}}a_{1}^{+}e^{-\alpha
ba_{1}^{+}a_{1}}=a_{1}^{+}(a_{2}^{+})^{\alpha }  \notag \\
Sa_{1}S^{-1} &=&e^{\alpha ba_{1}^{+}a_{1}}a_{1}e^{-\alpha
ba_{1}^{+}a_{1}}=a_{1}(a_{2}^{+})^{-\alpha }  \label{eq:16}
\end{eqnarray}%
and transformation of $a_{2}$ and $a_{2}^{+}$ as follows

\begin{eqnarray}
Sa_{2}^{+}S^{-1} &=&(a_{2}^{+})^{\alpha
a_{1}^{+}a_{1}}a_{2}^{+}(a_{2}^{+})^{-\alpha a_{1}^{+}a_{1}}=a_{2}^{+} 
\notag \\
Sa_{2}S^{-1} &=&(a_{2}^{+})^{\alpha a_{1}^{+}a_{1}}a_{2}(a_{2}^{+})^{-\alpha
a_{1}^{+}a_{1}}=a_{2}-\alpha a_{1}^{+}a_{1}(a_{2}^{+})^{-1}.  \label{eq:17}
\end{eqnarray}%
In a similar manner, from Schwinger representation to the Gelfand-Dyson
representation which is suitable to study QES systems, the generators of $%
su(2)$ and $su(1,1)$ algebra in the Schwinger representation can be
transformed by introducing the following operator:

\begin{equation}
T=a_{2}^{\alpha a_{1}^{+}a_{1}}  \label{eq:18}
\end{equation}%
The operator $T$ acts on the two-boson state as

\begin{equation}
T\left| n_{1},n_{2}\right\rangle =a_{2}^{\alpha n_{1}}\left|
n_{1},n_{2}\right\rangle =\sqrt{\frac{n_{2}!}{(n_{2}-\alpha n_{1})!}}\left|
n_{1},n_{2}-\alpha n_{1}\right\rangle .  \label{eq:19}
\end{equation}%
Since $a_{1}$ and $a_{2}$ commute, the transformation of $a_{1}$ and $%
a_{1}^{+}$ under $S$ can be obtained by letting $a_{2}=e^{c}$ with $%
[a_{1},c]=[a_{1}^{+},c]=0,$

\begin{eqnarray}
Ta_{1}^{+}T^{-1} &=&e^{\alpha ca_{1}^{+}a_{1}}a_{1}^{+}e^{-\alpha
ca_{1}^{+}a_{1}}=a_{1}^{+}(a_{2})^{\alpha }  \notag \\
Ta_{1}T^{-1} &=&e^{\alpha ca_{1}^{+}a_{1}}a_{1}e^{-\alpha
ca_{1}^{+}a_{1}}=a_{1}(a_{2})^{-\alpha }  \label{eq:20}
\end{eqnarray}%
The transformation of $a_{2}$ and $a_{2}^{+}$ are as follows:

\begin{eqnarray}
Ta_{2}^{+}T^{-1} &=&a_{2}^{\alpha a_{1}^{+}a_{1}}a_{2}^{+}a_{2}^{-\alpha
a_{1}^{+}a_{1}}=a_{2}^{+}+\alpha a_{1}^{+}a_{1}a_{2}^{-1}.  \notag \\
Ta_{2}T^{-1} &=&a_{2}^{\alpha a_{1}^{+}a_{1}}a_{2}a_{2}^{-\alpha
a_{1}^{+}a_{1}}=a_{2}  \label{eq:21}
\end{eqnarray}%
These two transformation leads to the two different $su(2)$ and two
different $su(1,1)$ realization in one dimensions and the corresponding
realizations are useful to study QES systems.

\section{\protect\bigskip Differential realizations of the su(2) and su(1,1)
algebras}

The realizations (\ref{eq:2}) and (\ref{eq:8}) can be transformed in the
form of the one dimensional differential equations in the Bargmann-Fock
space when the boson operators realized as%
\begin{equation}
a_{1}=\frac{d}{dx},\quad a_{1}^{+}=x.  \label{eq:22}
\end{equation}%
We can obtain two different differential realizations of the $su(2)$ and $%
su(1,1)$ algebras, depending on the choice of the $\alpha $, in the
equations (\ref{eq:14}) and (\ref{eq:17}).

\subsection{$su(2)$ Realization}

When the generators (\ref{eq:2}) of $su(2)$ algebra is transformed by using
the similarity transformation operator $S$ in the case of $\alpha =1,$ takes
the form:

\begin{eqnarray}
J_{+}^{\prime } &=&SJ_{+}S^{-1}=-(a_{1}^{+})^{2}a_{1}+N^{\prime }a_{1}^{+} 
\notag \\
J_{-}^{\prime } &=&SJ_{-}S^{-1}=a_{1}  \notag \\
J_{0}^{\prime } &=&SJ_{0}S^{-1}=a_{1}^{+}a_{1}-\frac{N^{\prime }}{2}  \notag
\\
N^{\prime } &=&SNS^{-1}=a_{2}^{+}a_{2}.  \label{eq:23}
\end{eqnarray}

These representations are called Gelfand-Dyson representation of the $su(2)$
algebra. Similarly we can easily obtain a second realization by using the
transformation operator $T$ and $\alpha =-1$. In this case the realization
of $su(2)$ is given by%
\begin{eqnarray}
J_{+}^{\prime } &=&TJ_{+}T^{-1}=a_{1}^{+}  \notag \\
J_{-}^{\prime } &=&TJ_{-}T^{-1}=-a_{1}^{+}(a_{1})^{2}+a_{1}(N^{\prime }-1) 
\notag \\
J_{0}^{\prime } &=&TJ_{0}T^{-1}=a_{1}^{+}a_{1}-\frac{N^{\prime }}{2}  \notag
\\
N^{\prime } &=&TNT^{-1}=a_{2}^{+}a_{2}.  \label{eq:24}
\end{eqnarray}

The difference between the Schwinger and Gelfand-Dyson representation is
that while in the first the total number of $a_{1}$ and $a_{2}$ bosons
characterize the the system, in the later it is only the number of $a_{2}$
bosons that characterize the system. According to (\ref{eq:8}) the
representation is characterized by a fixed number $2j$. Therefore in the
Gelfand-Dyson representation, the primed generators can be expressed in
terms of one boson operator $a_{1}$. According to the (\ref{eq:7}) takes the
values $N^{\prime }=2j$. The realizations (\ref{eq:23}) and (\ref{eq:24})are
the well known generators of the $sl(2,R)$ algebra, in the Bargmann-Fock
space, which play an important role in the quasi-exact solution of the Schr%
\"{o}dinger equation. The linear and bilinear combinations of the generators
form a second order differential equation and according to the Turbiner
theorem the linear and bilinear combinations of the generators $%
J_{i}^{\prime }$ $(i=+,-,0)$ are QES. The basis function of the primed
generators of the $su(2)$ algebra is the degree of polynomial of order $2j$,

\begin{equation}
P_{n}(x)=(x^{0},x^{1},\cdots ,x^{2j}).  \label{eq:25}
\end{equation}

\subsection{\protect\bigskip su(1,1) realizations}

By using the similar arguments given in previous subsection we can obtain
two different one dimensional differential realization for the $su(1,1)$
algebra in the Bargmann-Fock space. The generators of the $su(1,1)$ algebra
under the transformation of the $T$ when $\alpha =1$ takes the form%
\begin{eqnarray}
K_{+}^{\prime } &=&TK_{+}T^{-1}=(a_{1}^{+})^{2}a_{1}+(L^{\prime }+1)a_{1}^{+}
\notag \\
K_{-}^{\prime } &=&TK_{-}T^{-1}=a_{1}  \notag \\
K_{0}^{\prime } &=&TK_{0}T^{-1}=a_{1}^{+}a_{1}+\frac{L^{\prime }+1}{2} 
\notag \\
L^{\prime } &=&TLT^{-1}=a_{2}^{+}a_{2}.  \label{eq:26}
\end{eqnarray}%
The other realization can be obtained by transforming the generators with
the transformation operator S and choosing $\alpha =-1$:%
\begin{eqnarray}
K_{+}^{\prime } &=&TK_{+}T^{-1}=a_{1}^{+}  \notag \\
K_{-}^{\prime } &=&TK_{-}T^{-1}=a_{1}^{+}(a_{1})^{2}+(L^{\prime }+1)a_{1} 
\notag \\
K_{0}^{\prime } &=&TK_{0}T^{-1}=a_{1}^{+}a_{1}+\frac{L^{\prime }+1}{2} 
\notag \\
L^{\prime } &=&TLT^{-1}=a_{2}^{+}a_{2}.  \label{eq:27}
\end{eqnarray}%
The basis function of these generators are polynomials in $x$, in the
Bargmann-Fock space. In the representations (\ref{eq:26}) and (\ref{eq:27})
the operator $L^{\prime }$ characterize the system, $L^{\prime }=-2\ell -1.$

Consequently the (quasi)exact solution of the two-mode bosonic Hamiltonians
which include linear and bilinear combinations of the $su(2)$ or $su(1,1)$
algebra can be obtained by a suitable transformation.

\section{Applications}

In this section we discuss the applicability of the method to solve the
Hamiltonians of the various physical systems.

\subsection{Karassiov-Klimov Hamiltonian}

The method developed in this article can be used to obtain the solution of
the various two-boson Hamiltonians. Consider the following family of
Karassiov-Klimov\cite{kara2} Hamiltonians

\begin{equation}
H=\omega _{1}a_{1}^{+}a_{1}+\omega _{2}a_{2}^{+}a_{2}+\kappa
a_{1}^{+s}a_{2}^{r}+\bar{\kappa}a_{1}^{s}a_{2}^{+r}  \label{eq:28}
\end{equation}%
where $0<r<s$, $\omega _{1}$ and $\omega _{2}$ are frequencies of two
independent harmonic oscillator and $\kappa $ and $\bar{\kappa}$ are
coupling constants. The Hamiltonian was studied in the context of nonlinear
Lie algebras\cite{beck}, for $r=s=2$ and $r=1,\quad s=2$. Let us consider
the transformation of the Hamiltonian $H$ by the transformation operator $S$%
, choosing $\alpha =r/s$,%
\begin{equation}
H^{\prime }=SHS^{-1}=\omega _{1}a_{1}^{+}a_{1}+\omega _{2}(a_{2}^{+}a_{2}-%
\frac{r}{s}a_{1}^{+}a_{1})+\kappa a_{1}^{+s}(a_{2}^{+}a_{2}-\frac{r}{s}%
a_{1}^{+}a_{1})^{r}+\bar{\kappa}a_{1}^{s}  \label{eq:29}
\end{equation}%
Using the relation (\ref{eq:7}) and Bargmann-Fock space realizations (\ref%
{eq:22})of the $a_{1}$ and $a_{1}^{+}$ the Hamiltonian takes the form%
\begin{equation}
H^{\prime }=(\omega _{1}-\frac{r}{s})x\frac{d}{dx}+2j\omega _{2}+\kappa
x^{s}(2j-\frac{r}{s}x\frac{d}{dx})^{r}+\bar{\kappa}\frac{d^{s}}{dx^{s}}
\label{eq:30}
\end{equation}%
The eigenvalue problem can be written as%
\begin{equation}
H^{\prime }P_{n}(x)=EP_{n}(x).  \label{eq:31}
\end{equation}%
With the basis function (\ref{eq:25}) the eigenvalue equation (\ref{eq:31})
leads to the following recurrence relation%
\begin{equation}
(\omega _{1}-\frac{r}{s}-E+2j\omega _{2})P_{n}(E)+\kappa (2j-\frac{r}{s}%
n)^{r}P_{n+s}(E)+\bar{\kappa}\frac{n!}{(n-s)!}P_{n-s}(E)=0  \label{eq:32}
\end{equation}%
and the wave function of the Hamiltonian $H$ can be written as%
\begin{equation}
\psi (x)=S^{-1}\sum_{k=0}^{2j}P_{k}(E)x^{k}  \label{eq:33}
\end{equation}%
The wavefunction is itself the generating function of the energy
polynomials. The eigenvalues are then produced by the roots of such
polynomials. If the $E_{k}$ is a root of the polynomial $P_{k+1}(E)$, the
series (\ref{eq:22}) terminates at $k>2j\frac{s}{r}$ and $E_{k}$ belongs to
the spectrum of the corresponding Hamiltonian. The eigenvalues are then
obtained by finding the roots of such polynomials.

The Hamiltonian (\ref{eq:28}) have been considered in\cite{beck}, in the
context of the nonlinear algebras, for the specific values of $s=r=2$ and $%
s=2,r=1$. The Hamiltonian (\ref{eq:23}) can be expressed in terms of the
generators of $su(2)$ \ algebra when $s=r=2$

\begin{equation}
H=(\omega _{1}-\omega _{2})J_{0}+\kappa J_{+}^{2}+\bar{\kappa}J_{-}^{2}+%
\frac{1}{2}(\omega _{1}+\omega _{2})N  \label{eq:34}
\end{equation}%
and the transformed Hamiltonian can be cast into a differential operator

\begin{equation}
H=(\bar{\kappa}+\kappa x^{4})\frac{d^{2}}{dx^{2}}+x(\omega _{1}-\omega
_{2}+2\kappa (3-2j)x^{2})\frac{d}{dx}-2j(\omega _{1}+\kappa (1-2j)x^{2}),
\label{eq:35}
\end{equation}%
by using the realization given in (\ref{eq:23}). Our result is coincides
with the result given in\cite{beck}. The recurrence relations, with the
basis function (\ref{eq:25}) is given by

\begin{eqnarray}
\kappa (2j-n)(2j-n-1)P_{n+2}(\lambda )+\bar{\kappa}n(n-1)P_{n-2}(\lambda ) &&
\notag \\
+(\omega _{1}(n-2j)-\omega _{2}n-E)P_{n}(\lambda ) &=&0  \label{eq:36}
\end{eqnarray}

The Hamiltonians of the second and the third harmonic generation are the
special case of the Karassiov-Klimov Hamiltonian and they are given by 
\begin{subequations}
\begin{eqnarray}
H &=&\omega _{1}a_{1}^{+}a_{1}+\omega _{2}a_{2}^{+}a_{2}+\kappa
(a_{1}^{+2}a_{2}+a_{1}^{2}a_{2}^{+})  \label{eq:37a} \\
H &=&\omega _{1}a_{1}^{+}a_{1}+\omega _{2}a_{2}^{+}a_{2}+\kappa
(a_{1}^{+3}a_{2}+a_{1}^{3}a_{2}^{+})  \label{eq:37b}
\end{eqnarray}%
respectively. It is obvious that (\ref{eq:28}) can be put in the form of the
(\ref{eq:37a}-b) and they can be expressed as one variable differential
equations as in(\ref{eq:30}). Their eigenvalues and eigenfunctions can be
obtained by using the recurrence relation (\ref{eq:32}) and (\ref{eq:32}),
respectively.

\subsection{Quantal system under thermal effect}

The next example is the Hamiltonian of the quantal system under thermal
effect\cite{tsue1}. In order to investigate thermal effects in quantum
many-particle systems the Hamiltonian can be formulated by using two bosons
creation and annihilation operators. The Hamiltonians of the symmetric and
asymmetric rigid rotators have almost the same structure and their algebraic
structure is $su(1,1)$ algebra. The Hamiltonians of such systems can be
expressed as\cite{tsue2}: 
\end{subequations}
\begin{equation}
H=\hbar \omega (a_{1}^{+}a_{1}-a_{2}^{+}a_{2})-i\gamma \hbar
(a_{1}^{+}a_{2}^{+}-a_{1}a_{2})  \label{eq:38}
\end{equation}%
Using the realizations (\ref{eq:8}) of $su(1,1)$ algebra we can rewrite the
Hamiltonian:%
\begin{equation}
H=\hbar \omega L-i\gamma \hbar (K_{+}-K_{-})  \label{eq:39}
\end{equation}%
The Hamiltonian can be expressed as one dimensional differential equation by
using the transformation operators $S$ and $T$: 
\begin{subequations}
\begin{eqnarray}
H^{\prime } &=&SHS^{-1}=\hbar \omega (-2j-1)-i\gamma \hbar ((x^{2}-1)\frac{d%
}{dx}-2jx)  \label{eq:40a} \\
H^{\prime } &=&THT^{-1}=\hbar \omega (-2j-1)-i\gamma \hbar (x-x\frac{d^{2}}{%
dx^{2}}+2j\frac{d}{dx})  \label{eq:40b}
\end{eqnarray}%
respectively. According to the Turbiner theorem\cite{turb1} this equation is
exactly solvable.

\subsection{Two variable differential equations of the special functions}

Let us consider the following bosonic operator

\end{subequations}
\begin{equation}
H=\omega _{1}a_{1}^{+}a_{1}+\omega _{2}a_{2}^{+}a_{2}+\alpha
_{1}a_{1}^{+}a_{2}^{+}+\alpha _{2}a_{1}a_{2}-\frac{1}{2}a_{1}^{2}a_{2}^{2}.
\label{eq:41}
\end{equation}%
In terms of the generators of $su(1,1)$ the Hamiltonian can be expressed as

\begin{equation}
H=(\omega _{1}+\omega _{2})(K_{0}-\frac{1}{2})+\frac{1}{2}(\omega
_{1}-\omega _{2})L+\alpha _{1}J_{+}+\alpha _{2}J_{-}-\frac{1}{2}J_{-}^{2}
\label{eq:42}
\end{equation}%
The transformed Hamiltonian can be expressed as a differential operator

\begin{equation}
H=-\frac{1}{2}\frac{d^{2}}{dx^{2}}+(\alpha _{2}+x(\omega _{1}+\omega
_{2}+\alpha _{1}x))\frac{d}{dx}-(2k\alpha _{1}x+\omega _{1}(2k+1)),
\label{eq:43}
\end{equation}%
which can be written in the form of an eigenvalue equation

\begin{equation}
HR(x)=ER(x).  \label{eq:44}
\end{equation}%
The solution of (\ref{eq:43}) can be obtained by introducing the wave
function

\begin{equation}
R(x)=e^{-\int W(x)dx}\psi (x)  \label{eq:45}
\end{equation}%
where the weight function $W(x)$ is given by

\begin{equation}
W(x)=-\alpha _{2}-(\omega _{1}+\omega _{2})x-\alpha _{1}x^{2})  \label{eq:46}
\end{equation}%
With the wave function given in (\ref{eq:45}) the Hamiltonian (\ref{eq:43})
can be transformed in the form of Schr\"{o}dinger equation with the potential

\begin{eqnarray}
V(x) &=&\frac{1}{2}(\alpha _{2}^{2}-\omega _{1}(4k+3)-\omega _{2})+(\alpha
_{2}(\omega _{1}+\omega _{2})-\alpha _{1}(2k+1))x+  \notag \\
&&\frac{1}{2}(\alpha _{1}\alpha _{2}+(\omega _{1}+\omega
_{2})^{2})x^{2}+\alpha _{1}(\omega _{1}+\omega _{2})x^{3}+\frac{\alpha
_{1}^{2}}{2}x^{4}.  \label{eq:47}
\end{eqnarray}

Therefore the bosonic operator given by (\ref{eq:41}) is related to the
anharmonic oscillator potential. It is known that the Schr\"{o}dinger
equation with the potential given in(\ref{eq:47}) is QES. Let us turn our
attention to the Hamiltonian (\ref{eq:43}). When we set

\begin{equation}
\alpha _{1}=\alpha _{2}=0,\omega _{1}=\omega _{2}=1/2,E=\frac{1}{2}(n-2k-1)
\label{eq:48}
\end{equation}%
then the eigenvalue problem can be written as

\begin{equation}
\left( -\frac{1}{2}\frac{d^{2}}{dx^{2}}+x\frac{d}{dx}-\frac{n}{2}\right)
R(x)=0  \label{eq:49}
\end{equation}%
and the bosonic Hamiltonian takes the form

\begin{equation}
H=\frac{1}{2}(a_{1}^{+}a_{1}+a_{2}^{+}a_{2}-a_{1}^{2}a_{2}^{2}).
\label{eq:50}
\end{equation}%
It is obvious that differential equation of the two dimensional Hermite
polynomials in the Bargmann-Fock space can be expressed as:

\begin{equation}
\frac{1}{2}\left( -\frac{\partial ^{2}}{\partial x_{1}\partial x_{2}}+x_{1}%
\frac{\partial }{\partial x_{1}}+x_{2}\frac{\partial }{\partial x_{2}}%
\right) R(x_{1},x_{2})=0.  \label{eq:51}
\end{equation}

Our last example is the two-boson realization of the Schr\"{o}dinger
equation with the sextic harmonic oscillator potential. Let us consider the
following bosonic operator:

\begin{eqnarray}
L &=&(\omega _{1}+\alpha _{1})a_{1}^{+}a_{1}+(\omega _{1}-\alpha
_{1})a_{2}^{+}a_{2}+\alpha _{+}a_{1}^{+}a_{2}^{+}+  \label{eq:52} \\
&&(\alpha _{-}-\frac{1}{2})a_{1}a_{2}-\frac{1}{2}%
(a_{2}^{+}a_{1}a_{2}^{2}+a_{1}^{+}a_{2}a_{1}^{2})+\omega _{1}.  \notag
\end{eqnarray}%
In terms of the generators of the $su(1,1)$ algebra it can be written as

\begin{equation}
H=2\omega _{1}J_{0}+\alpha _{+}J_{+}+\alpha _{-}J_{-}+\alpha
_{1}N-J_{0}J_{-},  \label{eq:53}
\end{equation}%
which can be converted to the differential operator

\begin{equation}
H=-x\frac{d^{2}}{dx^{2}}+(\alpha _{-}+k+x(2\omega _{1}+\alpha _{+}x))\frac{d%
}{dx}-2k(\alpha _{+}+(\omega _{1}+\alpha _{+}x)).  \label{eq:54}
\end{equation}%
Let us change the variable and redefine the wave function for the eigenvalue
problem

\begin{equation}
H\psi (x)=E\psi (x),x=\left( \frac{z}{2}\right) ^{2},\psi (z)=e^{\int
W(z)dz}R(z)  \label{eq:55}
\end{equation}%
where the weight function $W(z)$ is given by

\begin{equation}
W(z)=\frac{\alpha _{+}}{16}z^{3}+\frac{\omega _{1}}{2}z+\frac{1+2\alpha
_{-}+2k}{2z}.  \label{eq:56}
\end{equation}%
Upon this substitution the potential term in the Hamiltonian can be written
as:

\begin{eqnarray}
V(x) &=&\alpha _{-}\omega _{1}-k(2\alpha _{1}+\omega _{1})+\frac{(1+2\alpha
_{-}+2k)(3+2\alpha _{-}+2k)}{4z^{2}}+  \notag \\
&&\frac{\alpha _{+}(\alpha _{-}-3k-1)+2\omega _{1}^{2}}{8}z^{2}+\frac{\alpha
_{+}\omega _{1}}{16}z^{4}+\left( \frac{\alpha _{+}}{16}\right) ^{2}z^{6}
\label{eq:57}
\end{eqnarray}%
This is the radial sextic oscillator potential studied in the literature.

\section{Conclusion}

In this paper we have discussed solution of the two-boson Hamiltonians and
applications in physics. It was shown that the solution of the two-boson
Hamiltonians can be transformed in the form of one-dimensional QES
differential equation by applying a suitable similarity transformation. This
transformation also leads to the connection between one and two-dimensional
special functions of the physics.

The method given here is useful to study nonlinear quantum optical systems.
The range of the Hamiltonian can be extended by the Bugoliov transformation
of the boson operators. It is expected that this work will leads to the
construction of the multi-boson QES Hamiltonians and their extensions to the
matrix Hamiltonians. We have been presented a first step towards an
extension of the quasi-exact solution of the multi-bosonic Hamiltonians.

\end{document}